\documentclass[reprint,amsmath,amssymb,aps]{revtex4-1}
\usepackage{graphicx,subfig}
\usepackage{bm}
\usepackage{color}
\usepackage[margin=2cm]{geometry}
\usepackage{epstopdf}

\begin{document}

\title{Prediction of a potential high-pressure structure of FeSiO$_3$}
\author{R. E. Cohen$^{1,2}$} \email[E-mail: ]{rcohen@carnegiescience.edu}
\author{Yangzheng Lin$^1$}
\affiliation{$^1$Geophysical Laboratory, Carnegie Institution of Washington, 5251 Broad Branch Road, NW, Washington DC 20015, USA}
\affiliation{$^2$Department of Earth Science, University College London, London, WC1E 6BT, UK}

\begin{abstract}
We predict a new candidate high-temperature high-pressure structure of FeSiO$_3$ with space-group symmetry Cmmm by applying an evolutionary algorithm within DFT+U that we call post-perovskite II (PPv-II). An exhaustive search found no other competitive candidate structures with ABO$_3$ composition. We compared the X-ray diffraction (XRD) pattern of FeSiO$_3$ PPv-II with experimental results of the recently reported ``H-phase'' of (Fe,Mg)SiO$_3$. The intensities and positions of two main X-ray diffraction peaks of PPv-II FeSiO$_3$ compare well with those of the H-phase. We also calculated the static equation of state, the enthalpy and the bulk modulus of the PPv-II phase and compared it with those of perovskite (Pv) and post-perovskite (PPv) phases of FeSiO$_3$. According to the static DFT+U computations the PPv-II phase of FeSiO$_3$ is less stable than Pv and PPv phases under lower mantle pressure conditions at T=0 K and has a higher volume. PPv-II may be entropically stabilized, and may be a stable phase in Earth’s lower mantle, coexisting with α-PbO$_2$ (Columbite-structure) silica and perovskite, or with magnesiowustite and/or ferropericlase, depending on bulk composition.
\end{abstract}


\maketitle


In 1987, Knittle and Jeanloz reported that silicate perovskite was stable throughout Earth's lower mantle \cite{Knittle87}, and for a long-time it was believed to be the last major phase change in Earth's mantle. Ten years ago, the post-perovskite phase was discovered, which explained many-features of Earth's D$^{\prime\prime}$ layer at the base of the mantle, and since has been widely believed to be the last mantle phase transition \cite{Hirose07}. Recently, a new phase has been reported formed by disproportionation of (Mg,Fe)SiO$_3$ perovskite \cite{Zhang14}.

We have used the evolutionary algorithm encoded using XtalOpt \cite{Lonie11,Xtalopt14} with the quantum espresso PWSCF code \cite{Giannozzi09,Pwscf14} and GGA(PBE)+U by searching for the lowest enthalpy phases of FeSiO$_3$ at 100 GPa with 10, 15, 20, and 30 atoms/primitive cell. The reported cell-parameters for the H-phase were used at initial guess, and each structure was optimized at a constant pressure of 100 GPa. 

After calculations for several thousands of FeSiO$_3$ structures, we selected some with the lowest enthalpies for further studies. The symmetry of a selected structure was refined by FINDSYM \cite{Stokes05,ISOtropy14} with tolerance no higher than 0.01. After careful study on the selected structures, we obtained three phases of FeSiO$_3$ that have the lowest enthalpies at 100 GPa. Perovskite and post-perovskite are two among the three. Another low enthalpy phase, referred to as the post-perovskite II (PPv-II) here, was not discovered before.

The PPv-II structure is C-centered orthorhombic, with a pseudohexagonal primitive cell (Fig. \ref{fig:Figures01} a and c, Table \ref{tab:Tables01}). This structure consists of silicon layers stacking along the a-direction and intercalated Fe ions, which is the same as in the structure of post-perovskite FeSiO$_3$. In the silicon layer of post-perovskite FeSiO$_3$, one interval line of silicon atoms and one interval line of Fe atoms move about half of its lattice constant along the a-direction (corresponding to c-direction in the structure of PPv-II) to form post-perovskite II. Accompanying the movement of silicon and Fe lines, some oxygen atoms become bonded to three silicon atoms, while other oxygen atoms become bonded only to one silicon atom (Fig. \ref{fig:Figures01}). PPv-II looks like PPv compressed along the silicon surfaces and also along the Fe surfaces while pulled in the vertical direction, which is demonstrated by the lattice lengths in Table \ref{tab:Tables01}. In PPv-II, the average Si-O bond distance is smaller than that of PPv, while the average Fe-O bond distance is larger {(Table \ref{tab:Tables01}).

\begin{widetext}

\begin{figure}[h]
\subfloat[PPv-II]{\includegraphics[height=2.7in]{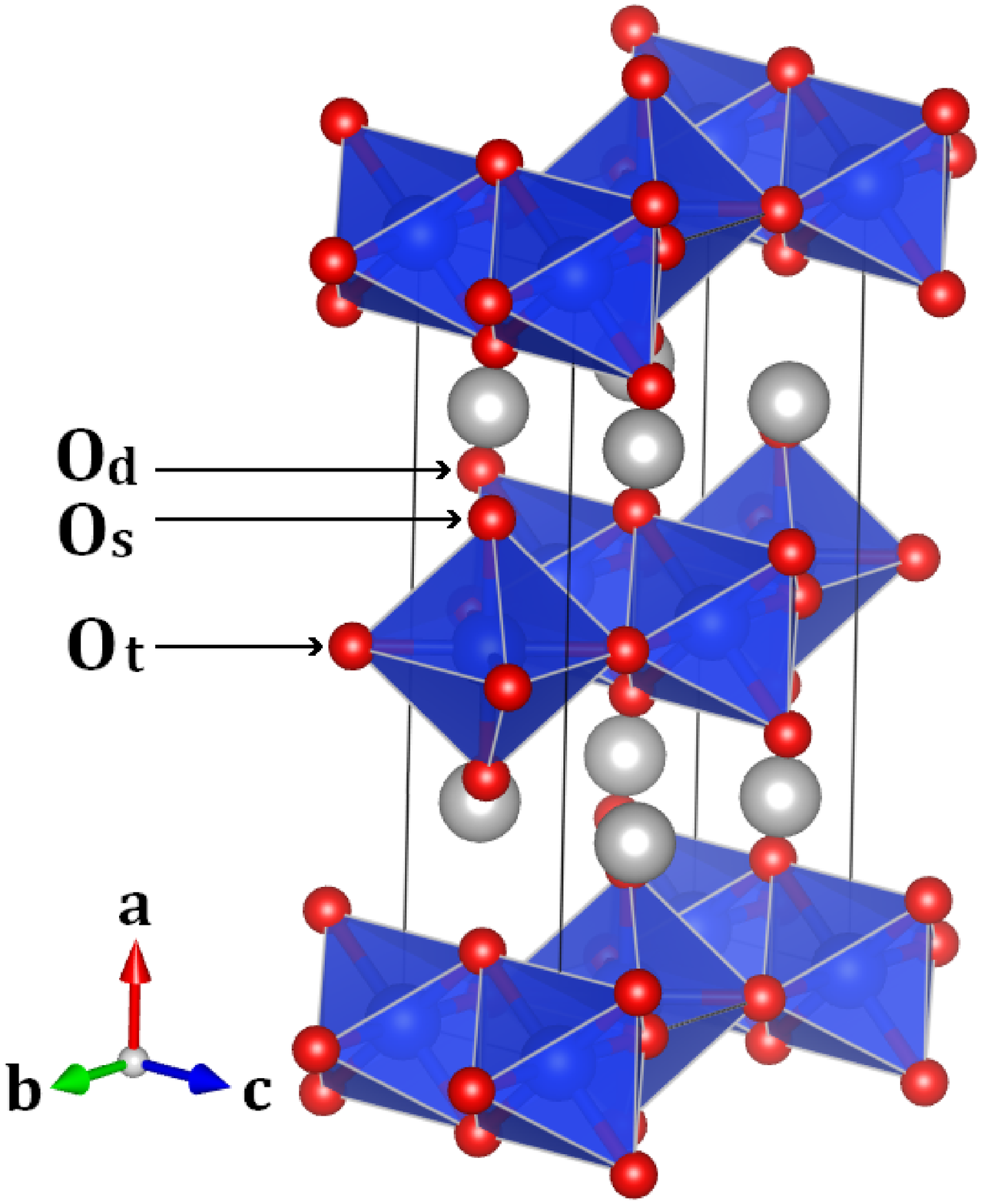}}
\subfloat[PPv]{\includegraphics[height=2.7in]{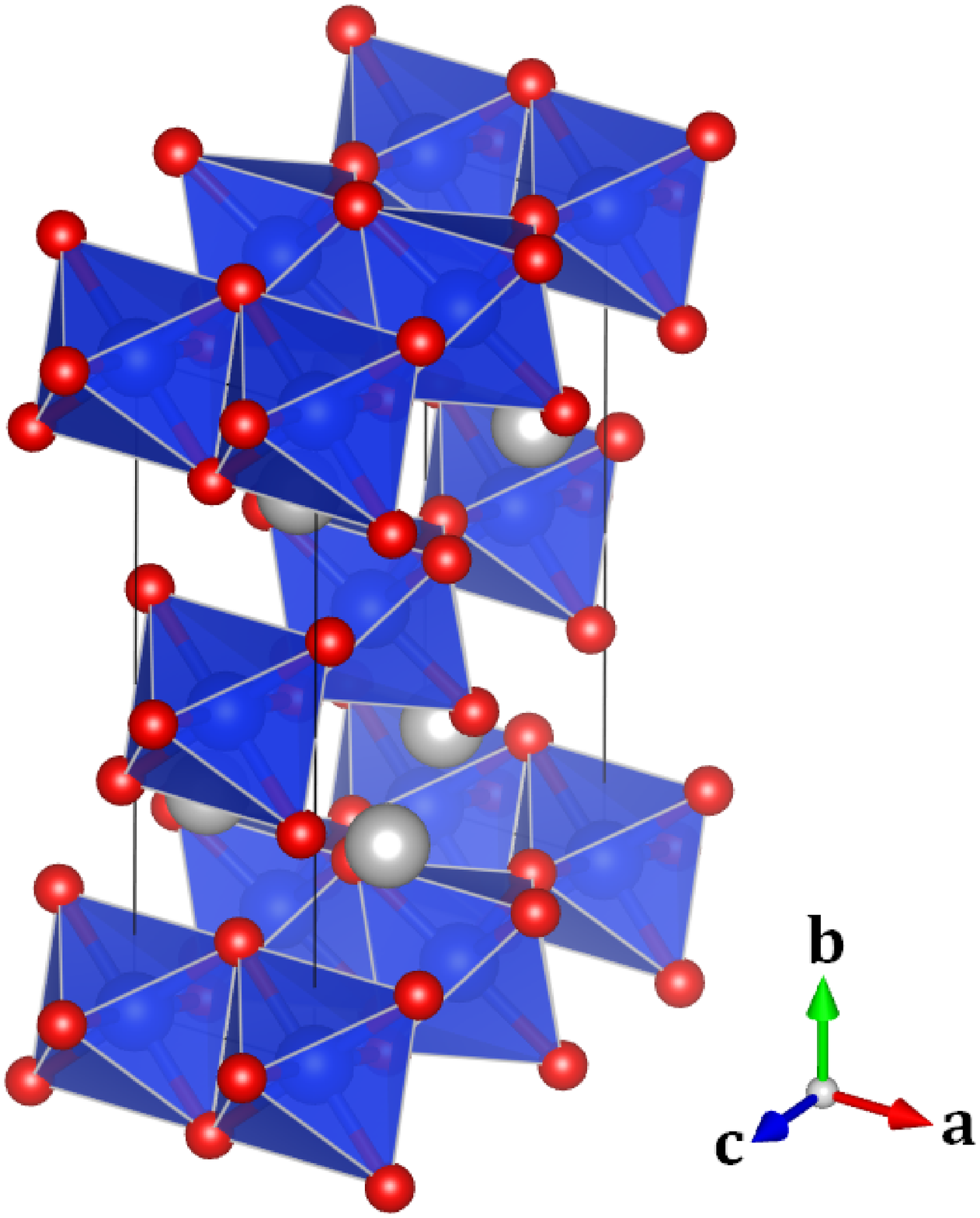}}
\space{}
\subfloat[PPv-II from c direction]{\includegraphics[height=1.4in]{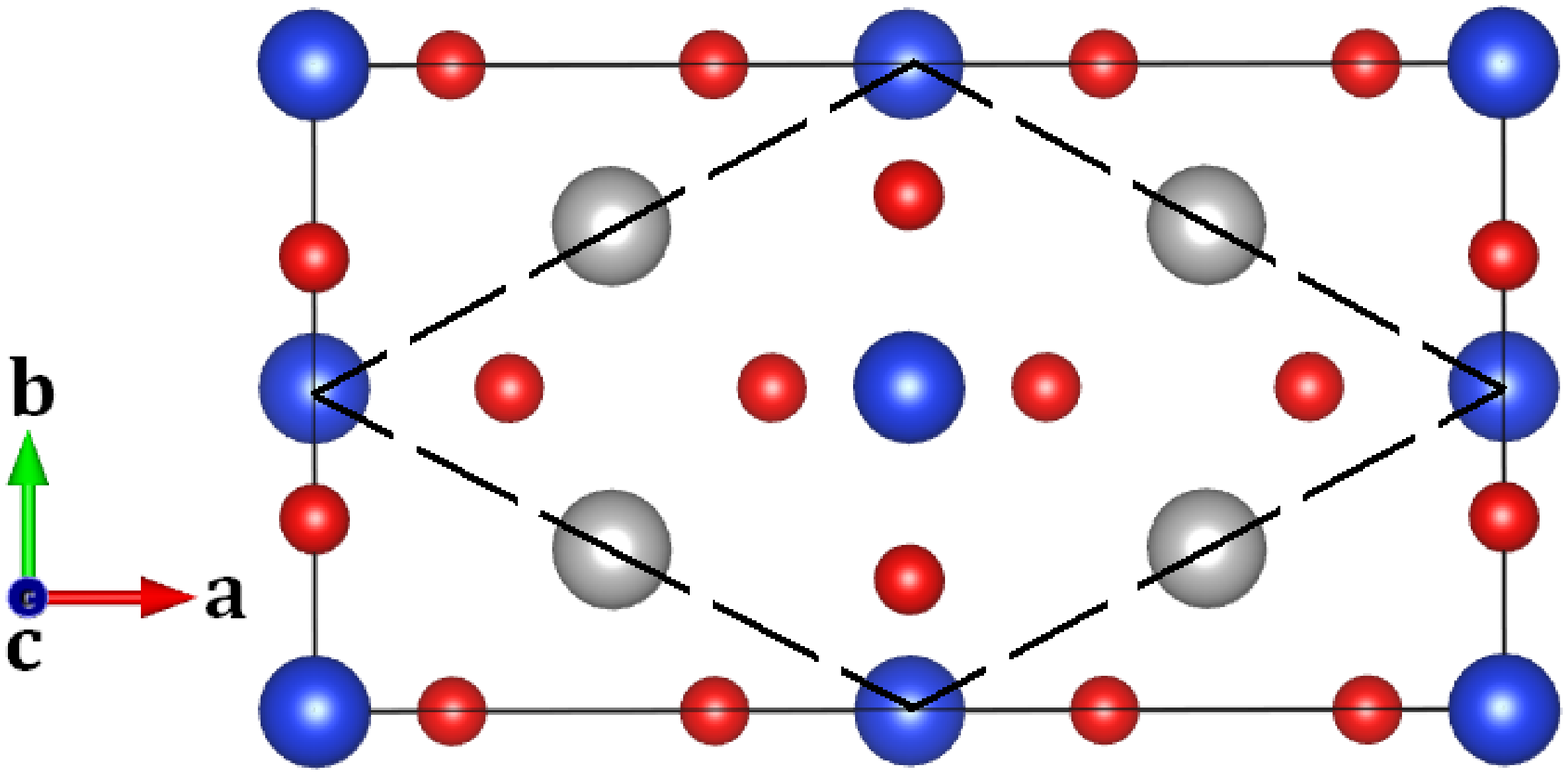}}
\subfloat[PPv from a direction]{\includegraphics[height=1.4in]{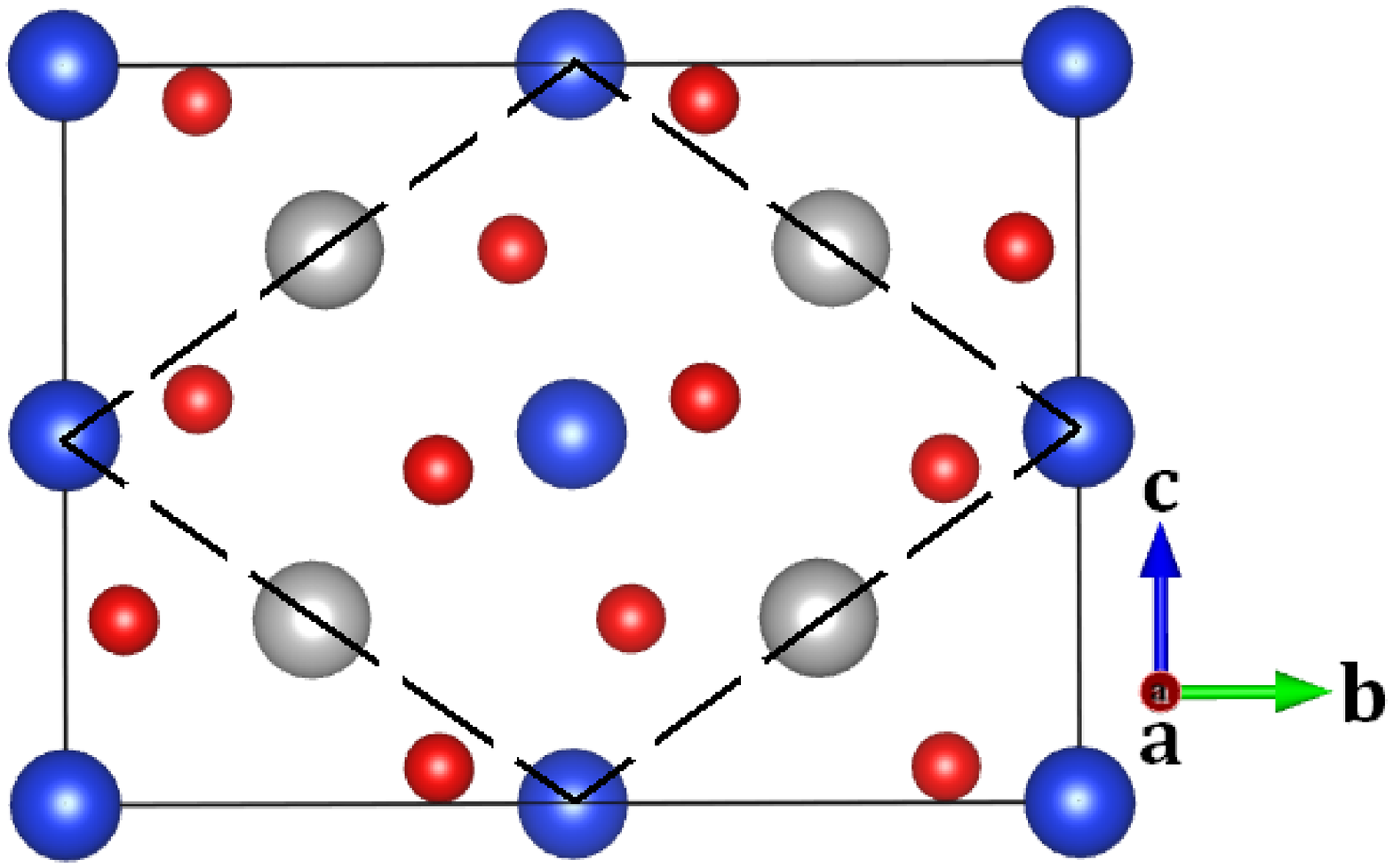}}
\caption{The atomic structure of FeSiO$_3$ post-perovskite II phase (a) and comparison with post-peroviskte FeSiO$_3$ (b) in one unit cell. (c) is the view of PPv-II from c direction and (d) is the view of PPv from a direction. The gray, blue, and red spheres represent Fe, Si and O atoms respectively. The octahedra represent Si with six-fold oxygen coordination. In PPv-II, there are three kinds of O atoms who are in different sharing situations. O$_s$ only bonds to one silicon atom, O$_d$ bonds to two silicon atoms, and O$_t$ is sharing by three silicon atoms in the crystal. While all oxygen atoms in PPv are sharing by two silicon atoms.}
\label{fig:Figures01}
\end{figure}

\begin{table}[h]
\centering
\caption{Structure information of PPv-II and PPv phases of FeSiO$_3$ at 100 GPa from GGA+U}
\label{tab:Tables01}
\begin{tabular}{c@{ }c@{ }c@{ }c@{ }c@{ }c@{ }c@{ }c@{ }c@{ }}
\hline
 Phase          & \multicolumn{4}{c}{PPv-II}        & \multicolumn{4}{c}{PPv} \\
\hline
 Crystal system & \multicolumn{4}{c}{Orthorhombic}  & \multicolumn{4}{c}{Orthorhombic} \\
 Space group    & \multicolumn{4}{c}{Cmmm (No. 65)} & \multicolumn{4}{c}{Cmcm (No. 63)} \\
\hline
 \multicolumn{9}{l}{Cell parameters} \\
   &   & a(\AA)   & b(\AA)  & c(\AA)   &   & a(\AA)   & b(\AA)  & c(\AA) \\
   &   & 10.082   & 5.478   & 2.495    &   & 2.508    & 8.614   & 6.283 \\
   &   & $\alpha$($^{\circ}$) & $\beta$($^{\circ}$) & $\gamma$($^{\circ}$) &   & $\alpha$($^{\circ}$) & $\beta$($^{\circ}$) & $\gamma$($^{\circ}$) \\
   &   & 90 & 90 & 90 &   & 90 & 90 & 90 \\
\hline
 \multicolumn{9}{l}{Atomic coordinates} \\
    & {Label*} & $x$ & $y$ & $z$ & {Label*} & $x$ & $y$ & $z$ \\
Fe  & 4 e & 0.25  & 0.25  & 0   & 4 c & 0   & 0.743 & 0.25 \\
Si1 & 2 a & 0     & 0     & 0   & 4 a & 0   & 0     & 0    \\
Si2 & 2 c & {0.5}   & {0}     & 0.5 &     &     &       &      \\
O1  & 4 i & 0     & 0.702 & 0   & 4 c & 0   & 0.058 & 0.25 \\
O2  & 4 h & 0.336 & 0     & 0.5 & 8 f & 0   & 0.368 & 0.048 \\
O3  & 4 h & 0.115 & 0     & 0.5 &     &     &       &       \\
\hline
 \multicolumn{9}{l}{Interatomic distances (\AA)} \\
 Si1-O & \multicolumn{4}{c}{1.630($\times$2), 1.705($\times$4)} & \multicolumn{4}{c}{1.648($\times$2), 1.718($\times$4)} \\
 Si2-O & \multicolumn{4}{c}{1.652($\times$2), 1.669($\times$4)} & \multicolumn{4}{c}{ } \\
Average Si-O & \multicolumn{4}{c}{1.672} & \multicolumn{4}{c}{1.695} \\
 Fe-O & \multicolumn{4}{c}{2.042($\times$4), 2.297($\times$4}) & \multicolumn{4}{c}{2.030($\times$2), 2.083($\times$4), 2.107($\times$2)} \\
Average Fe-O & \multicolumn{4}{c}{2.170} & \multicolumn{4}{c}{2.076} \\
\hline
 \multicolumn{9}{l}{*The label letters are Wyckoff labels and the number before each letter is the corresponding multiplicity.} \\
\end{tabular}
\end{table}

\end{widetext}

We calculated the X-ray diffraction patterns of PPv FeSiO$_3$, PPv-II FeSiO$_3$ and Pv MgSiO$_3$ under experimental pressures and wavelengths (Fig. \ref{fig:Figures02}). According to Ref. \cite{Zhang14}, the experimental powder is composed mainly of Pv MgSiO$_3$ and H-phase in Fig. \ref{fig:Figures02}A, and the experimental powder is nearly pure H-phase in Fig. \ref{fig:Figures02}B. H110 and H101 are the two main peaks of H-phase based on these experimental XRD patterns. Surprisingly PPv-II FeSiO$_3$ also has two main peaks PPv-II400 and PPv-II220 whose positions are very close to the peak positions of H110 and H101 (Fig. \ref{fig:Figures02}) in spite of that the lattice lengths of pseudohexagonal primitive cell of PPv-II are different from those of the hexagonal H-phase in Ref. \cite{Zhang14} (Table \ref{tab:Tables02}).

\begin{table}[h]
\centering
\caption{Lattice parameters comparison between the primitive cell of PPv-II and the H-phase in Ref. \cite{Zhang14} at 100 GPa.}
\label{tab:Tables02}
\begin{tabular}{c@{ }c@{ }c@{ }c@{ }c@{ }c@{ }c@{ }}
\hline
 & a(\AA) & b(\AA) & c(\AA) & $\alpha$($^{\circ}$) & $\beta$($^{\circ}$) & $\gamma$($^{\circ}$) \\ 
\hline
 H-phase \cite{Zhang14} & 5.096 & 5.096 & 2.862 & 90 & 90 & 120 \\
 PPv-II Primitive cell  & 5.737 & 5.737 & 2.495 & 90 & 90 & 122.96 \\
\hline
\end{tabular}
\end{table}

\begin{figure}[h]
\centering
\includegraphics[width=3.2in]{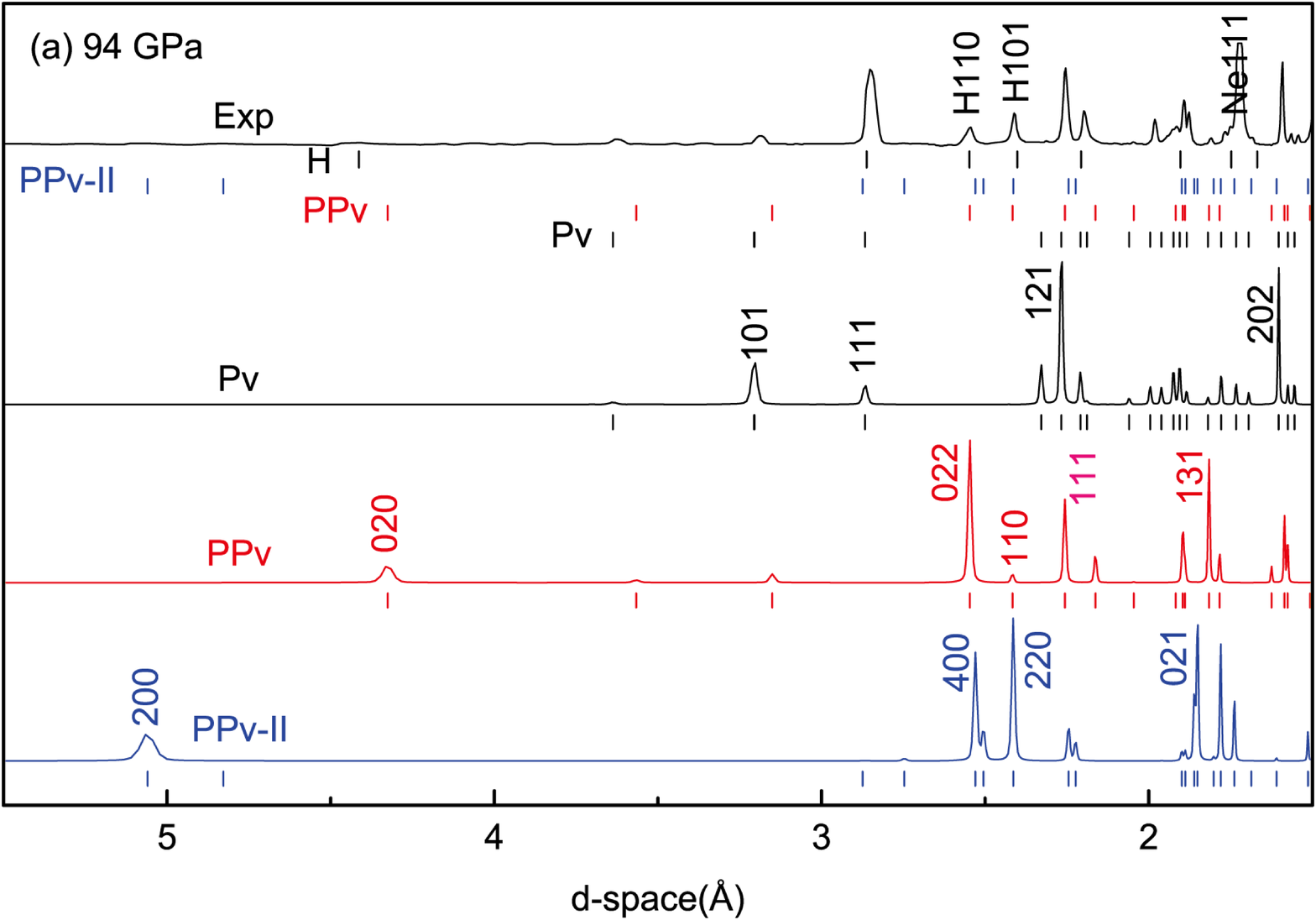}
\includegraphics[width=3.2in]{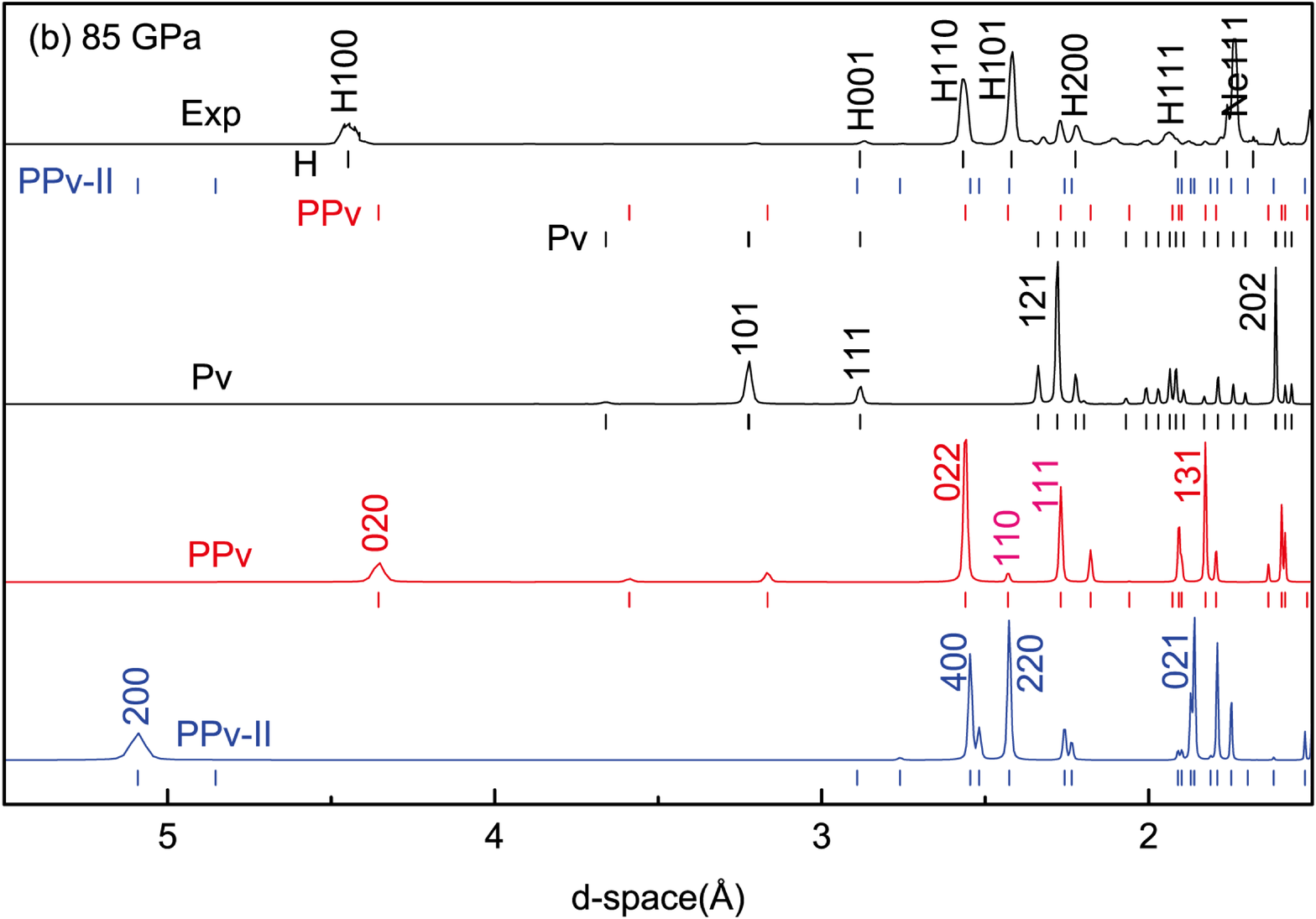}
\caption{Calculated X-ray diffraction patterns of perovskite (Pv) MgSiO$_3$, post-perovskite (PPv) FeSiO$_3$ and post-perovskite II (PPv-II) FeSiO$_3$ crystals at 94 and 85 GPa and their comparison with experiments. (A) Experimental pattern was read from Fig. 1C in Ref. \cite{Zhang14} and (B) Experimental patter was read from Fig. 4(a) in Ref. \cite{Zhang14}. The horizontal axis indicates d-space here instead of 2$\theta$ in the original figures.}
\label{fig:Figures02}
\end{figure}

Note that the experimental intensities are not likely to be reliable as they come from averaged diffraction of small single-crystallites (i.e. diffraction spots rather than rings). The differences in peak positions of PPv-II from experiment is caused by the difference in composition - ours is for pure FeSiO$_3$ , whereas the experimental pattern contains some Mg. Note that PPv also has two peaks at the peak positions of H110 and H101. Zhang et al. claim that PPv is ruled out by their data: ``In addition to Pv, another set of peaks not corresponding to any previously known phases appeared with particularly conspicuous peaks at 2.55 and 2.40 \AA  (marked H110 and H101, respectively). These two peaks are close to the diagnostic PPv peaks near 2.5 and 2.4 \AA  (PPv 022 and 110, respectively), but the high-quality XRD pattern clearly rejects the possibility of PPv" \cite{Zhang14}. The best diagnostic peak for PPv-II would be the large d-spacing 200 peak, but this weak small 2$\theta$ peak may be difficult to observe.

We studied the equations of state (EOS) of Pv, PPv, and PPv-II phases of FeSiO$_3$ based on DFT+U static calculations. All of our calculations used projector augmented-wave (PAWs) generated with Perdew-Burke-Ernzerhof (PBE) exchange-correlation functional in the Quantum Espresso library for Fe, Si and O atoms. We used 20-atoms cells, 4$\times$4$\times$4 k-points mesh and 80 Ry cut-off energy for all Pv, PPv, and PPv-II crystals so that the enthalpy difference between FeSiO$_3$ phases converged within 0.01eV/FeSiO$_3$. We used U$=$6.0 which is the same value as used in the structure searchings. All the phases of FeSiO$_3$ are in high-spin antiferromagnetic states in our calculations for they have lower total energy than the corresponding ferromagnetic and low-spin states.

For each phase of FeSiO$_3$, we calculated the static total energies at eight volumes (Fig. \ref{fig:Figures03}). The eight volumes correspond to eight pressures of -10, 0, 25, 50, 75, 100, 125 and 150 GPa. The EOS parameters were fitted to the E(V) data with a Vinet equation of state \cite{Vinet87,Cohen00}. At 100 GPa, the bulk modulus of PPv-II is smaller than those of Pv and PPv and the one formula volume of PPv-II is larger (Table \ref{tab:Tables03}).

\begin{figure}[h]
\centering
\includegraphics[width=3.2in]{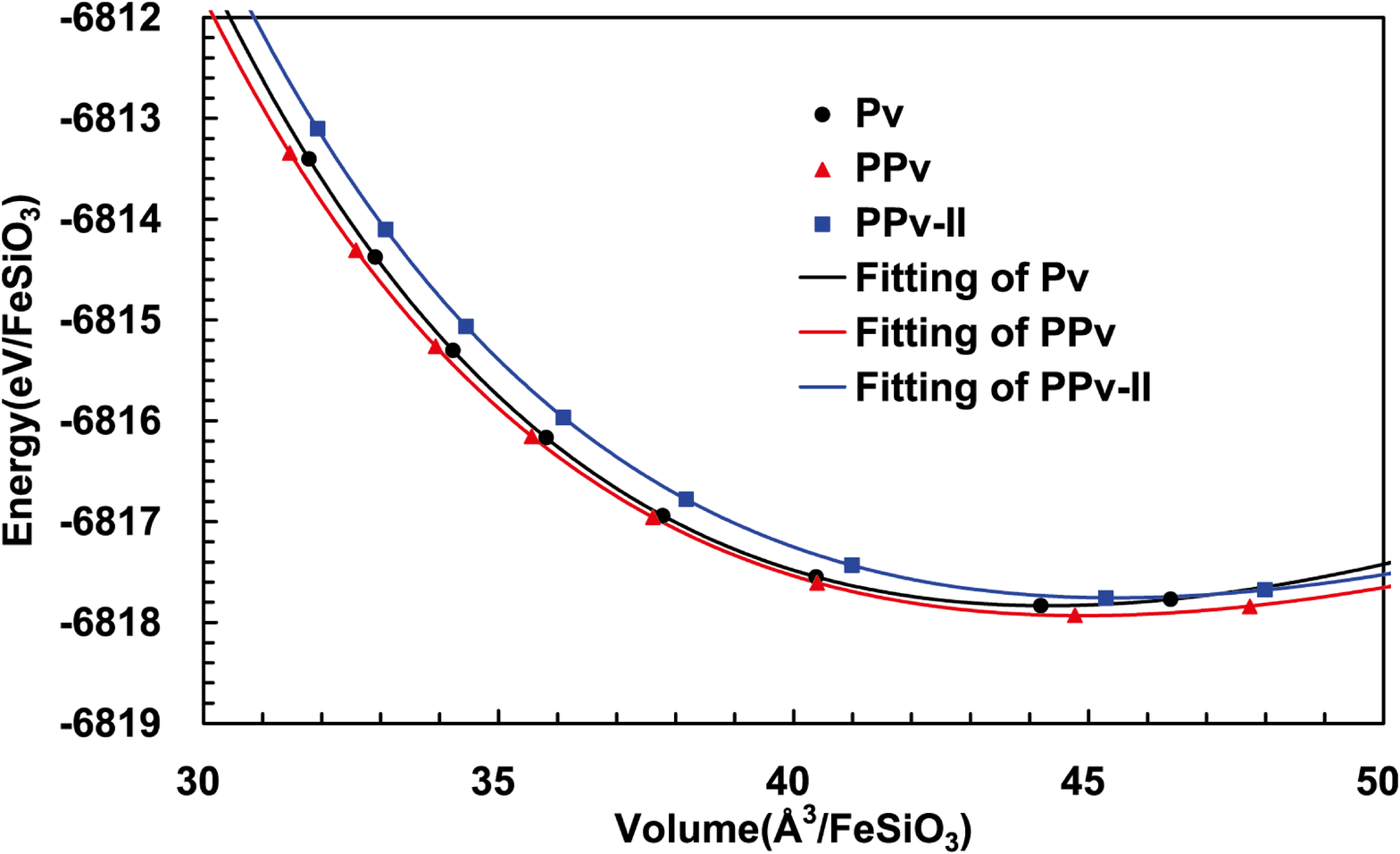}
\caption{Static energies of FeSiO$_3$ Pv, PPv and PPv-II phases using GGA+U. The solid curves were fitted using the Vinet equation of state \cite{Vinet87,Cohen00}.}
\label{fig:Figures03}
\end{figure}

\begin{table}[h]
\centering
\caption{Parameters of the Vinet equation of state fitted to the GGA+U static total energy ($V$ is volume, $K$ is bulk modulus, $K^{\prime}=\text{d}K/\text{d}P$ where $P$ is pressure, the subscripts 0 and 100 indicate the values at 0 and 100 GPa respectively.)}
\label{tab:Tables03}
\begin{tabular}{c@{ }c@{ }c@{ }c@{ }}
\hline
  & $V_0$($V_{100}$,\AA$^3$/FeSiO$_3$) & $K_0$($K_{100}$,GPa) &  $K_0^{\prime}$($K_{100}^{\prime}$) \\
\hline
 Pv     & 44.31(34.27) & 225(597) & 4.42(3.34)  \\
 PPv    & 44.90(33.98) & 189(579) & 4.73(3.47) \\
 PPv-II & 45.45(34.49) & 195(580) & 4.67(3.44) \\
\hline
\end{tabular}
\end{table}

Pv and PPv phases of FeSiO$_3$ are more stable than the PPv-II phase at high pressures in the lower mantle (Fig. \ref{fig:Figures04}). We find that the orthorhombic structure (Cmmm) of PPv-II is unstable at low pressure and it has a displacive phase transition to monoclinic symmetry (C2/m) (Fig. \ref{fig:Figures04}). The monoclinic structure of PPv-II phase of FeSiO$_3$ is more stable than Pv at pressures lower than 10.8 GPa and more stable than PPv at 4.5 GPa.

\begin{figure}[h]
\centering
\includegraphics[width=3.2in]{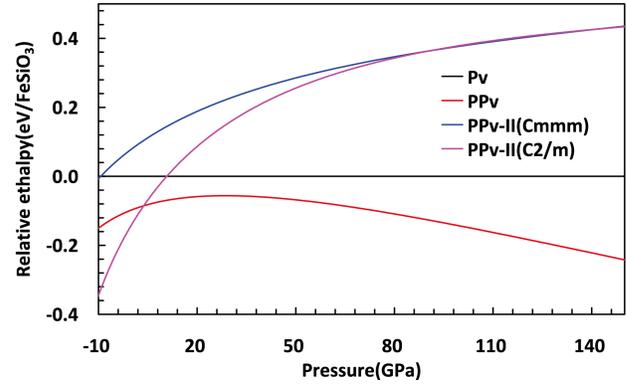}
\caption{The relative enthalpy of Pv, PPv and PPv-II phases of FeSiO$_3$ from GGA+U (Enthalpy of Pv FeSiO$_3$ is zero as reference).}
\label{fig:Figures04}
\end{figure}

It is intriguing to consider more complex phases that have a mixture or super lattice of the PPv Si-O chains and the PPv-II Si-O chains. These could either form as a stable structure, or due to kinetically hindered solid-state reactions. It also seems possible that the Fe atoms could move in their planes at high temperatures, as in host-guest structures \cite{Degtyareva07}.

We have discovered a post-perovskite II phase of FeSiO$_3$ using crystal structure searches based on enthalpy from DFT calculations at 100 GPa. The crystal of PPv-II is orthorhombic. The structure of PPv-II can be formed from PPv by moving interval lines of silicon and iron atoms half the lattice constant along their surfaces. Based on the enthalpy from the Vinet equation of state regressed from static total energies of 8 volumes, the PPv-II phase of FeSiO$_3$ is less stable than its Pv and PPv phases at lower mantle pressure conditions. PPv-II has slightly larger volume than Pv and PPv, and high temperature is probably more propitious to PPv-II than Pv and PPv. Two main peaks in the XRD pattern of PPv-II are in good agreement with the experimental peaks of the H-phase both in their positions and relative intensities.

\section{Acknowledgments}
\label{acknowledgments}
This work is supported by National Science Foundation grants DMS-1025370 and EAR-1214807. R. E. Cohen was supported by the Carnegie Institution and by the European Research Council advanced grant ToMCaT. We thank Li Zhang for helpful discussions.

\clearpage
\bibliographystyle{apsrev4-1}
\bibliography{FeSiO3-PPv-II}{}

\end{document}